# Evaluation of CuO$_2$ plane hole doping in YBa$_2$Cu$_3$O$_{6+x}$ single crystals

Ruixing Liang, D. A. Bonn, W. N. Hardy
*Department of Physics and Astronomy, University of British Columbia, 6224 Agricultural Road,
Vancouver, B. C. Canada V6T 1Z1*

The electron hole concentration $p$ in the CuO$_2$ planes of YBa$_2$Cu$_3$O$_{6+x}$ (YBCO) has been evaluated using the unit cell length in the $c$-direction together with a simple estimation for $p$ in the chain oxygen ordered phases. The empirical relationship between $p$ and $c$ obtained allows calculation of $p$ from the $c$-axis lattice parameter for the entire doping range of YBCO. It is also suggested that the empirical parabola describing the relationship between $T_c$ and $p$ in La$_{2-\delta}$Sr$_\delta$CuO$_4$ is reasonably correct for YBCO except for the region around $p$ =1/8, where the $T_c$ of YBCO is suppressed by as much as 17 K. A correction of the parabola for the $p$ = 1/8 effect is presented.

The parent compounds of all high $T_c$ cuprates are antiferromagnetic insulators where the CuO$_2$ plane Cu $d_{x^2-y^2}$ orbital is exactly half-filled. When holes are doped into the planes the antiferromagnetic ordering weakens and eventually gives way to high $T_c$ superconductivity. The doping $p$, the number of holes per copper atom in the CuO$_2$ planes, is a parameter that appears in all theories of high $T_c$ superconductivity and is the single most important parameter that determines the properties of high $T_c$ cuprates. The doping dependence of various physical properties provides key insights into the problem of high $T_c$ superconductivity and has been a major focus of high $T_c$ research.

The determination of the doping dependence requires accurate values of $p$. This is not a problem for some cuprates such as La$_{2-\delta}$Sr$_\delta$CuO$_4$ (LSCO), where $p$ is simply equal to the substitution concentration $\delta$. It has been found that the superconducting transition temperature $T_c$ and doping $p$ in LSCO have a parabolic relationship[1-3]:

$$1- T_c/ T_{c,\max} = 82.6\ (p-0.16)^2 \qquad (1)$$

where $T_{c,\max}$ is the maximum $T_c$ for the compound. However, for most other cuprates $p$ is not obvious from their chemical formula and, in fact, is very difficult to determine. YBa$_2$Cu$_3$O$_{6+x}$ (YBCO), the most widely investigated cuprate, has two different Cu-sites, the chain site Cu(1) and plane site Cu(2), and it is not clear how holes are distributed between the two Cu-sites. For oxygen content $x$ between 0.35 and 0.55 the Hall coefficient $R_H$ gives very reasonable values of $p$.[4] For other values of oxygen content, however, $R_H$ is strongly temperature dependent. In any case it is the hole density in the CuO$_2$ planes alone, that is of greatest interest. Conduction of the chains will contribute to $R_H$ and complicate the situation. Presumably, at low oxygen content the chains conduct poorly, which may account for the observed behavior of $R_H$.

Due to the difficulty in measuring $p$, it is now a widely adopted practice to calculate $p$ for YBCO (and other cuprates) from the superconducting $T_c$ using the empirical Eq. (1). However, Eq. (1) was obtained from data for LSCO. Although it may also be correct for Bi and Tl based cuprates[1], it has not yet been established to be correct for YBCO. Furthermore, when Eq. (1) was obtained, the data near $p$ = 1/8 were excluded for the regression. At this doping, $T_c$ is suppressed by the tendency of charge stripe formation and does not follow Eq. (1).[1-3] Therefore, before applying Eq. (1) to YBCO one ought to ask the question if, and to what extent, the $T_c$ of YBCO is also suppressed around 1/8 doping. Furthermore, in the very underdoped region where $T_c$ = 0, Eq. (1) is certainly not applicable and an alternative method is needed to evaluate $p$.

In YBCO, each Cu(2) atom in the CuO$_2$ planes has an apical oxygen O(4) above it. This oxygen moves toward the plane when $p$ increases, that is, when positive charge in the plane increases. Properties related to the Cu(2)-O(4) bonding, such as the Raman frequency of O(4) A$_g$ mode[5-7] and the Cu(2)-O(4) bond length[8,9] have been found to be sensitive to the change in oxygen content $x$ that is closely related to hole doping $p$. However, neither Raman frequencies nor bond lengths are conveniently measurable. Another quantity sensitive to the change in oxygen content is the $c$-direction unit cell length which is a sum of bond lengths, $c = 2d_{\text{Cu}(1)-\text{O}(4)} + d_{\text{Cu}(2)-\text{Cu}(2)} + 2d_{\text{Cu}(2)-\text{O}(4)}$. The change in $c$ is mainly caused by the change in $d_{\text{Cu}(2)-\text{O}(4)}$ because it is much more sensitive to change in the oxygen content than other bond lengths. The lattice parameter $c$ can be easily measured to high precision by diffraction techniques and there has been considerable work[7-12] on the relationship between $c$ and the oxygen content of YBCO. However, by examining the data for our single crystals we have found that the dependences of $p$ and $c$ on the oxygen content are not unique: both $p$ and $c$ depend not only on the oxygen content but also on the degree of oxygen *ordering* in the Cu(1)-O chains. We argue that there is however a unique relationship between $p$ and $c$.

The lattice constant $c$ was studied in high purity (99.995%) YBCO crystals grown in barium zirconate crucibles.[13] Crystals with oxygen content $x$ = 0 were prepared by annealing the crystals at 660°C under oxygen partial pressure 1 × 10$^{-5}$ Pa. Crystals with $x$ between 0.09 and 0.99 were prepared using procedures reported elsewhere.[14,15] The oxygen content was determined by weight analysis[15] to a precision of ±0.002. The chain oxygen ordered ortho-II, ortho-VIII and ortho-III phases were prepared with $x$ = 0.50, 0.67, and 0.75, respectively, by annealing the crystals at 10°C below the corresponding ordering temperature[16], 105°C, 40°C, and 75°C, respectively. The superconducting $T_c$ of the crystals was determined by measuring the in-field cooling magnetization at 1 Oe applied parallel to the $c$-axis, with $T_c$ being chosen as the mid-point of the transition. The lattice parameter $c$ was determined from (00$l$) X-ray diffraction lines measured at room temperature (22°C) using CuKα radiations. To minimize systematic errors only diffraction lines with 2θ greater than 100° were used. The data was extrapolated to 2θ = 180° to yield $c$.

The Cu(1)-O chain segments promote holes to the planes only when they are at least the critical length of three Cu atoms (two oxygen atom in between).[17] When the oxygen content is increased from $x$ = 0, where YBCO is tetragonal, initially hole doping increases slowly since oxygen ions in the chain sites are far apart and only a small fraction of them are in Cu(1)-O chain segments at or longer than the critical length. Therefore $c$ should change slowly with oxygen content near $x$ = 0 if the change in $c$ is dominated by the change in doping in the planes. We have confirmed this is indeed the case: when the oxygen content $x$ is increased from 0 to 0.10, $c$ changes from 1.18447 nm to 1.18403 nm or by only 0.037%. In comparison, when $x$ changes from 0.35 to 0.45, $c$ changes from 1.17833 to 1.17631 nm or by 0.17%.

Fig. 1(a) shows the superconducting transition temperature, $T_c$, of YBCO as a function of oxygen content. This curve, with a tendency towards a plateau near $x$ = 0.66 and a maximum near $x$ = 0.9 is well known. The reason for the $T_c \cong$ 60K quasi-plateau between $x$ = 0.5 and 0.75 is not yet well understood. One explanation is that when the oxygen content is increased from $x$ = 0.50, where YBCO forms the ortho-II phase with alternating full (means the chain oxygen site is fully occupied) and empty (means the chain oxygen site is empty) chains, additional oxygen ions fill the empty chains and make a relatively small contribution to hole doping, since they are far apart from each other and only a small fraction of them are in Cu(1)-O chain segments at or longer than the critical length.[17] However, this model fails to explain why the quasi-plateau extends smoothly beyond the ortho-II phase region and into regions of ortho-VIII (ordering in sequence of full-empty-full-full-empty-full-full-empty) and ortho-III (full-full-empty) phases.[16]

A major complication of Fig. 1 is that the dependence of $T_c$ on oxygen content is not unique because the doping $p$ depends also on the oxygen order in the Cu(1)-O chain layer. This is particularly obvious for $x$ < 0.50 as shown in Fig. (a), where the lower dashed line shows the $T_c$ for the ortho-I phase which has equal average oxygen occupancy for *every* Cu(1)-O chain, and the upper dashed line shows the $T_c$ for the ortho-II phase which has alternating empty and partially filled chains. When $T_c$ is plotted against $c$, as shown in Fig. 1(b), the data fall onto a single curve. This indicates that $c$ is more uniquely related to the doping than the oxygen content. A striking feature of the plot is that the curve is very similar in shape to the curve of $T_c$ vs. oxygen content and also has the $T_c \cong$ 60 K quasi-plateau. It seems unlikely that the slope of $c$ versus $p$ is suddenly changing in this regime, so Fig. 1(b) suggests that the 60 K "plateau" in $T_c$ is an *intrinsic* doping effect, and not caused by oxygen ions entering the empty chain sites. That is, it is $T_c$ itself that plateaus for some other reason, not the doping.

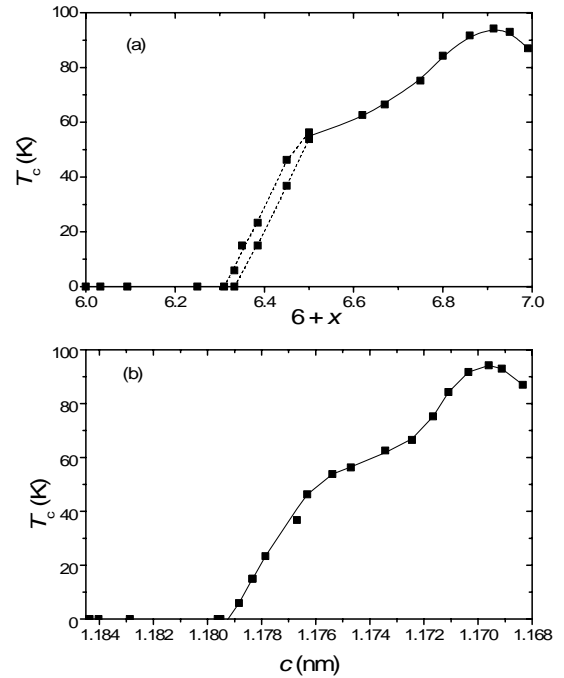

Fig. 1. Superconducting $T_c$ of YBCO as functions of (a) oxygen content and (b) $c$-axis lattice parameter. The shape of the two curves is quite similar.

As a first step in determining the relationship between $p$ and $c$, the doping $p$ was calculated from the superconducting $T_c$ using Eq. (1) and $T_{c,max}$ = 94.3 K for high purity crystals, and plotted against the lattice parameter $c$, as shown in Fig. 2. The data between $p$ = 0.05 and 0.09, corresponding to $T_c$ between 0 and 52 K, show linearity between $p$ and $c$. In particular, the linearity extrapolates to the data point for $p$ = 0. This suggests that Eq. (1) generates reasonable values of $p$ for YBCO in the low-doping region, as is also confirmed by the measurements of Hall coefficient.[4] However, in the median doping region between $p = 0.09$ and 0.15, corresponding to $T_c$ between 52 and 90 K, the data unexpectedly falls below the extrapolation from the low doping region. Since this median doping region is around $p = 1/8$, a natural explanation for this deviation is that $T_c$ is depressed by a tendency towards stripe formation, which Eq. (1) does not account for, causing the $p$ value calculated using Eq. (1) to be lower than it actually is. It is worth mentioning here that for YBCO $p = 1/8$ is a characteristic point where a minimum in the coherence length[18] and a change in dependence of $T_c$ on the superfluid density[19] have been observed.

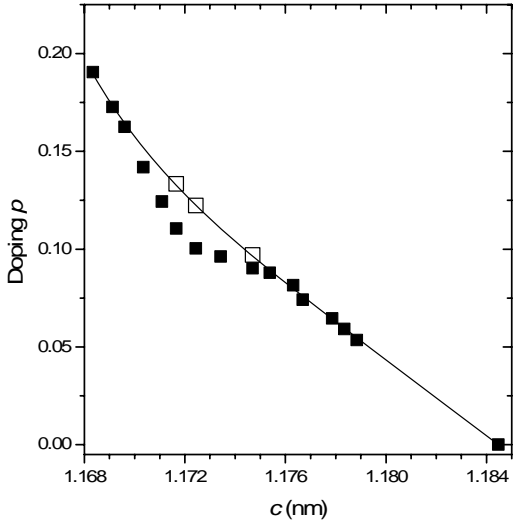

Fig. 2. The doping $p$ of YBCO as a function of the lattice parameter $c$. Solid squares represent $p$ values obtained using Eq. (1). Open squares are $p$ values estimated for the chain oxygen ordered phases (see text). The solid line is the fit to the data of $p > 0.15$ and $p < 0.09$ obtained using Eq. (1) in conjunction with values estimated for the ordered phases.

We now make use of the fact that the chain oxygen ordered phases ortho-II, ortho-VIII and ortho-III happen to fall in the median doping region around $p = 1/8$. Using simple arguments we can determine $p$ for the ordered phases relative to that for $x = 1$. In particular we assume that each oxygen ion in a full Cu(1)-O chain makes the same contribution to the doping regardless of whether the neighbouring chain is empty or full, so that $p$ in these phases is essentially proportional to the fraction of the Cu(1)-O chains that are full. This assumption is reasonable if the coupling between the chains is very weak. It has been reported that the chains show very strong one-dimensional characteristics.[4] For $x = 1$ where every chain is full, we obtained $p = 0.194$ by inserting $T_c = 85.3$ K (see Fig. 1(a)) into Eq. (1), assuming Eq. (1) is correct in the slightly overdoped region. Therefore for ortho-II YBCO ($x = 0.50$) where half of the chains are full, the doping $p = 0.097$ is half of that for $x = 1$. Next, the ortho-III phase has two third of the chains full.

However, the best ortho-III ordering occurs at $x = 0.75$,[16] which is higher than the theoretical oxygen content $x = 0.667$, indicating that the "empty" chains have a finite oxygen occupancy $v = 0.22$. Such an "empty" chain also makes a small contribution to the doping. Based on the calculation by Zaanen et al[17] it should be a good approximation that a chain segment of $n$ oxygen ions makes the same contribution to the doping as $n - 1$ oxygen ions in full chains. Thus the doping contribution by an "empty" chain is $v^2$ times the contribution by a full chain if oxygen ions distribute randomly in the "empty" chain, which should be the case when $v$ is small. Therefore, the doping in the ortho-III YBCO is $p = 0.194 \times (2/3 + v^2/3) = 0.132$. Similarly, we can calculate $p$ for the ortho-VIII (five eighths of the chains are full and $x = 0.67$). The $p$ values estimated for these ordered phases are plotted in Fig. 2 as open squares. We see that the data fall exactly on the extension of the linear low doping region and smoothly connect to data points for $p > 0.15$. This suggests the $p$ values estimated by the simple method are very reasonable.

Combination of the data by Eq. (1) for the low doing region ($p < 0.09$), optimal doping region ($p > 0.15$) and the data estimated for the chain oxygen ordered phases suggests a smooth relationship between $p$ and $c$. Fitting these data yields:

$$p = 11.491y + 5.17 \times 10^9 y^6 \qquad (2)$$

where $y = 1 - c/c_0$ and $c_0 = 1.18447$ nm (at 22ºC) is the $c$-axis lattice parameter at zero doping ($x = 0$). The second term is negligible for the low doping region ($p < 0.1$). This empirical formula enables calculation of $p$ from the easily measurable lattice parameter $c$ over the entire doping range for YBCO.

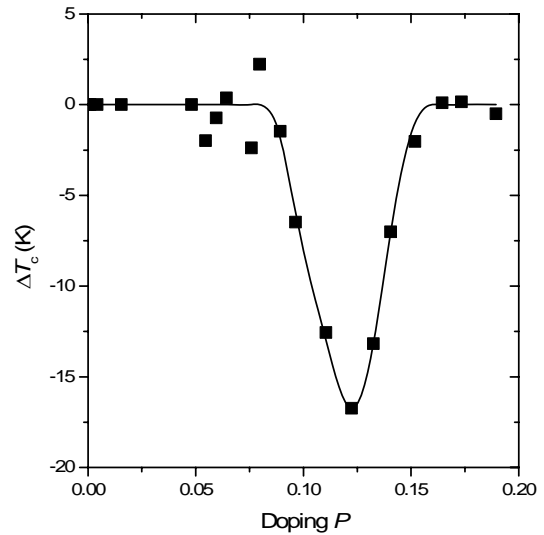

Fig. 3. $\Delta T_c$, the difference between actual $T_c$ and $T_c$ calculated by Eq. (1) using the $p$ values obtained from Eq. (2).

The line is a guide to the eye. The $T_c$ suppression reaches a maximum very close to $p = 1/8 = 0.125$.

Eq. (2), together with the actual measured values of critical temperature $T_{c,act}$ can be used to generate a relationship between $T_{c,act}$ and $p$. We compare this to the commonly used parabolic relationship by using Eq. (1) to calculate a value $T_{c,cal}$. This comparison is made in Fig. 3 which shows a plot of the difference $\Delta T_c = T_{c,act} - T_{c,cal}$ as a function of doping $p$. As is seen, the suppression of $T_c$ reaches a maximum very close to $p = 1/8$. The maximum suppression, 17 K or about 20%, is comparable to that for LSCO.[2] A much more pronounced $T_c$ suppression was observed in $La_{2-\delta}Ba_\delta CuO_4$ (LBCO).[20] The fact that $T_c$ is suppressed at the same doping as has been seen in the LSCO system lends further confidence to our determination of $p$ from the $c$-axis lattice constant.

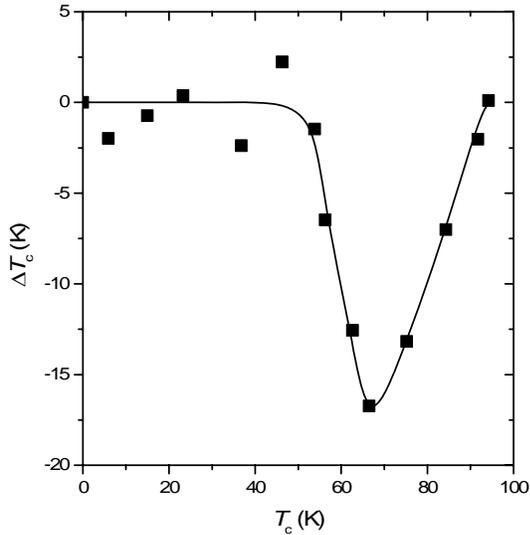

Fig. 4. Plot of $p = 1/8$ related $T_c$ suppression vs. $T_c$. The line is a guide to the eye. This plot is for correction of the empirical Eq. (1).

When $T_c$ is finite it is very convenient to calculate $p$ from $T_c$ using Eq. (1). Based on the analysis presented in this paper, we provide a correction to Eq. (1) in Fig. 4 where the $T_c$ suppression data is plotted as a function of $T_c$. The plot allows correcting for the 1/8 doping related $T_c$ suppression before applying Eq. (1) to calculate $p$.

In summary, we have evaluated the doping $p$ in YBCO using the $c$-axis lattice parameter and a simple method for estimating $p$ in chain oxygen ordered phases. The relationship between $p$ and $c$, Eq. (2), allows calculation of $p$ from $c$ for the entire doping region of YBCO, including the region where $T_c = 0$. The relationship calibrates well against the characteristic doping $p = 1/8$. Our data also suggest that Eq. (1) obtained from data for LSCO is reasonably correct for YBCO in the low doping region ($p < 0.09$) and near the optimal doping ($p > 0.15$). This, together with the early study on Bi and Tl based cuprates,[1] suggests that the parabolic Eq. (1) is universal for hole doped cuprates except near $p = 1/8$ where Eq. (1) is incorrect due to $T_c$ suppression by the tendency of charge stripe formation. The magnitude of the $T_c$ suppression is compound specific. For YBCO, our data suggest the suppression occurs between $p = 0.09$ and 0.15 with a maximum suppression of 17 K, or about 20%, at $p = 1/8$. The correction to Eq. (1) for the effect of $p = 1/8$ is generated, which allows convenient use of Eq. (1) when $T_c$ is finite.

This work was funded by the Natural Science and Engineering Research Council of Canada and the Canadian Institute for Advanced Research. DAB is grateful for NSF support of the Aspen Centre for Theoretical Physics.